\documentclass{cimento}

\usepackage{graphicx}

\newcommand{\met}  {\mbox{$\not\!\!E_T$}}
\newcommand{\rar}  {\rightarrow}

\title{Physics beyond the single top quark observation}

\author{A.P.~Heinson\\for the {D\O} Collaboration}

\instlist{\inst{}University of California, Riverside,
                  CA 92521-0413, USA}

\PACSes{\PACSit{14.65.Ha}{Top quarks}}

\begin{document}

\maketitle


\vspace{-0.1in}

\begin{abstract}
In March 2009, the {D\O} Collaboration first observed the electroweak
production of single top quarks at 5$\sigma$ significance. We measured
the cross section for the combined s-channel and t-channel production
modes, and set a lower limit on the CKM matrix element
$|V_{tb}|$. Since then, we have used the same dataset to measure the
t-channel production mode independently, the combined cross section in
the hadronically-decaying tau lepton final state, and the width and
lifetime of the top quark, and we have set upper limits on
contributions from anomalous flavor-changing neutral currents. This
paper describes these new measurements, as presented at the 3rd
International Workshop on Top Quark Physics, held in Brugge, Belgium,
May~31 -- June~4, 2010.
\end{abstract}


\vspace{-0.3in}

\section{Introduction}

The focus of this workshop, the top quark, is arguably the most
interesting known particle of the standard model. Since the top quark
decays before it can hadronize, many of its properties are directly
accessible to experimenters, unlike for other quarks. But the main
reason for our interest is that since the top quark is by far the
heaviest of the elementary particles ($m_t = 173.3 \pm
1.1$~GeV~\cite{topmass}), it may have properties and couplings not
predicted by the standard model. Detailed study of its production and
decay modes could thus provide a window to new physics.

\begin{figure}[!h!tbp]
\begin{center}
\includegraphics[width=1.75in]{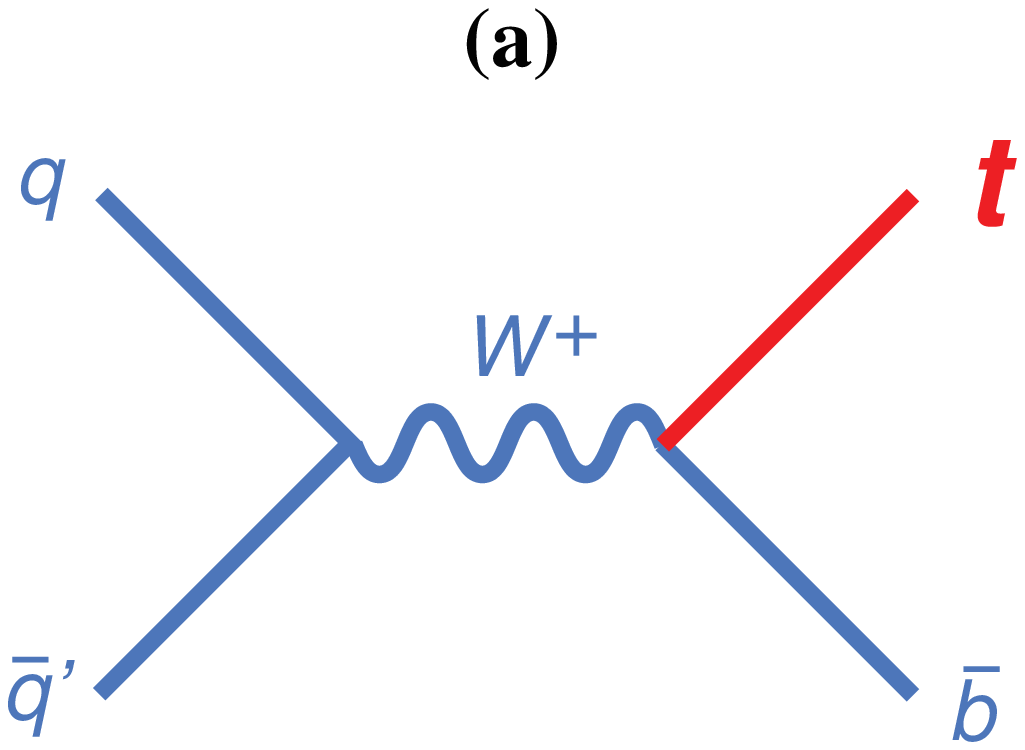}\hspace{0.25in}
\includegraphics[width=1.75in]{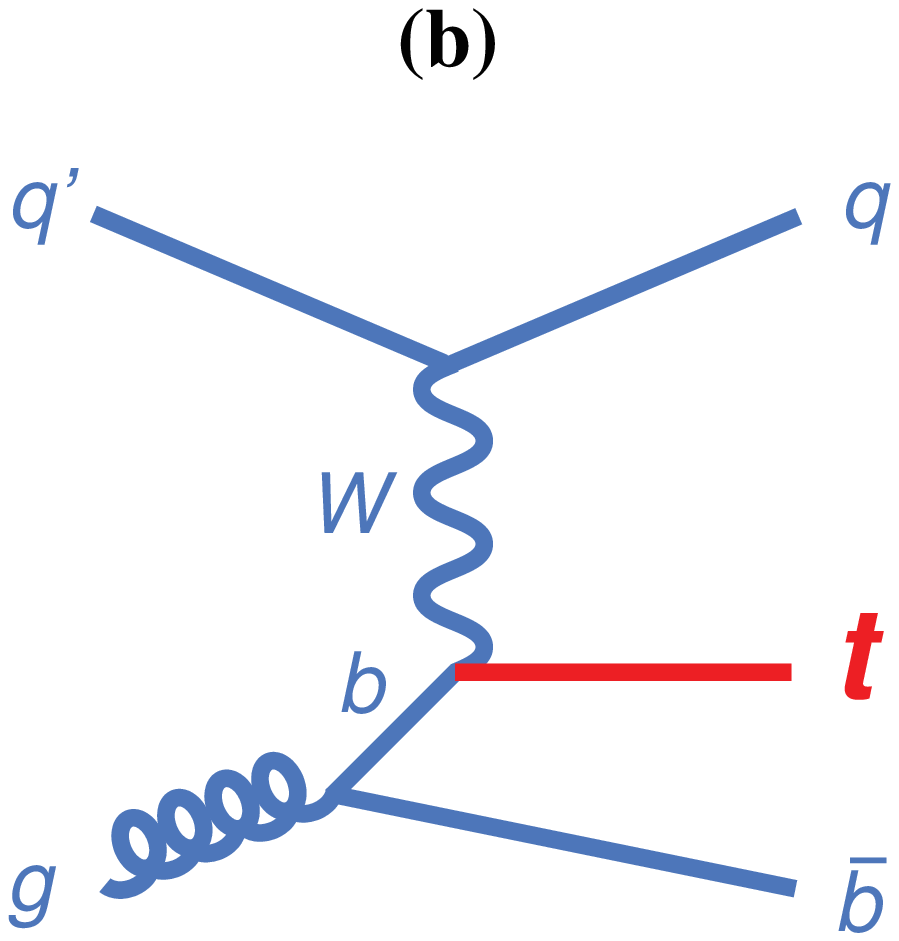}
\caption{Representative leading-order Feynman diagrams for the
electroweak production of single top quarks at the Tevatron: (a)
s-channel $tb$ production, and (b) t-channel $tqb$ production.}
\label{feynman}
\end{center}
\end{figure}

Top quarks are mostly produced in particle-antiparticle pairs via the
strong interaction from a very high energy virtual gluon. The cross
section at the Tevatron 1.96~TeV proton-antiproton collider is about
7.5~pb~\cite{ttbarxsec}. They can also be produced singly from a
highly energetic virtual $W$~boson via the electroweak
interaction~\cite{singletoptheory}. If the $W$~boson is in the
\mbox{s-channel}, then it decays to a top quark and an antibottom
quark or an antitop quark and a bottom quark, known together as
``$tb$'' production, with a cross section of $\approx
1$~pb~\cite{singletopxsec1,singletopxsec2}. If the $W$~boson is in the
t-channel or u-channel, then it fuses with a bottom quark to produce
the top quark, and there are an accompanying light quark and
antibottom quark; this is known collectively as ``$tqb$'' or t-channel
production, with a $\approx 2$~pb cross
section~\cite{singletopxsec1,singletopxsec2}. A third mode, $gb\rar
tW$, is negligible at the Tevatron. The labels ``s,''``t,'' and ``u''
are Mandelstam variables that describe the four-momenta of the
interactions~\cite{mandelstam}. The main tree-level Feynman diagrams
for single top quark production at the Tevatron are shown in
Fig.~\ref{feynman}.

\clearpage

In the standard model there are only three generations of quarks, and
in this situation, the CKM quark mixing matrix element $V_{tb}$ is
constrained indirectly to be very close to one. Therefore, the top
quark decays almost 100\% of the time to a $W$~boson and a bottom
quark. In the analyses presented here, we select events where the
$W$~boson decays leptonically to an electron, muon, or tau, and a
neutrino.


\section{Single top quark observation}

The new analyses presented here are based on the events selected for
the observation measurement of single top quark
production~\cite{dzerosingletopobs}. Many details of the measurement
can be found in a recent long review paper~\cite{heinsonmpla} and
proceedings paper~\cite{heinsonblois}. The review paper also includes
details of the simultaneous observation measurement by the CDF
collaboration~\cite{cdfsingletopobs} and the combination of {D\O} and
CDF results~\cite{combination}.

A brief summary of the event selection and analysis is given
here. From 2.3~fb$^{-1}$ of data, 1.2 billion events were selected
that passed any reasonable trigger and contained an electron or
muon. We chose from this dataset 4,519 events that passed offline
electron or muon identification, had large missing transverse energy
(to indicate the presence of a neutrino), and that had two, three, or
four jets, where one or two of the jets were identified as having
originated from the decay of a $b$ hadron. We built a detailed model
of the expected background processes and verified that it reproduced
all aspects of the data in regions not expected to contain signal. We
used three multivariate discrimination tools trained on an independent
subset of the signal and background models to separate signal from
background, and used a combination of them in a binned likelihood
calculation to extract the single top quark production cross section
and a limit on $|V_{tb}|$ at $m_t = 170$~GeV:

\vspace{-0.2in}
\begin{eqnarray*}
&\sigma(p\bar{p}\rar tb+X,tqb+X) = 3.94 \pm 0.88 {\rm ~pb} \\
&0.78 < |V_{tb}| \le 1 {\rm ~~~~at~95\%~C.L.}
\end{eqnarray*}
\vspace{-0.2in}

\noindent The NLO theory prediction at this mass is
$3.46{\rm ~pb} \pm 5\%$~\cite{singletopxsec2}.

Of the 22\% total uncertainty on the cross section measurement, 18\%
came from the data statistics of the measurement, and we inferred that
the systematic uncertainties contributed the remaining 13\% (when
combined in quadrature). The main components of the systematic
uncertainty came from the modeling of the $b$~jet identification in
Monte Carlo events, the modeling of the jet energy scale, and the
fraction of $W$+heavy flavor jets in the $W$+jets background model.


\section{Single top quark production in the t-channel}

For the observation analysis, s-channel and t-channel single top
processes were combined as signal in order to maximize the chance to
reach 5$\sigma$ significance. However, it is more interesting to
measure the cross section of each process separately, since they can
be affected by physics beyond the standard model in different
ways. Nonstandard couplings would change the kinematics and angular
distributions for example. The existence of a fourth generation of
quarks would affect $|V_{tb}|$ and hence the $b$-tagged fraction of
jets. Resonances such as a heavy $W^{\prime}$~boson, charged Higgs
boson, Kaluza-Klein excited $W$~boson, technipion, or a top-flavor $X$
particle would change the s-channel cross section only. And
flavor-changing neutral currents would affect both channels, although
dominantly the s-channel.

To measure the t-channel cross section independently of any assumption
about the ratio of s-channel to t-channel cross sections (note, the SM
ratio was assumed in the observation analysis), we trained new
discriminants on the background model and data from the observation
analysis, using only t-channel single top quark Monte Carlo events as
signal. Repeating all subsequent steps of the analysis as before, we
measured the following cross sections for $m_t = 170$~GeV:

\vspace{-0.2in}
\begin{eqnarray*}
\sigma(p\bar{p}\rar tqb+X) &=& 3.14^{+0.94}_{-0.80} {\rm ~pb}\\
\sigma(p\bar{p}\rar tb+X)  &=& 1.05 \pm 0.81 {\rm ~pb}\\
{\rm \mbox{t-channel}~significance} &=& 4.8\sigma.
\end{eqnarray*}
\vspace{-0.2in}

\begin{figure}[!h!tbp]
\begin{center}
\includegraphics[width=2.5in]{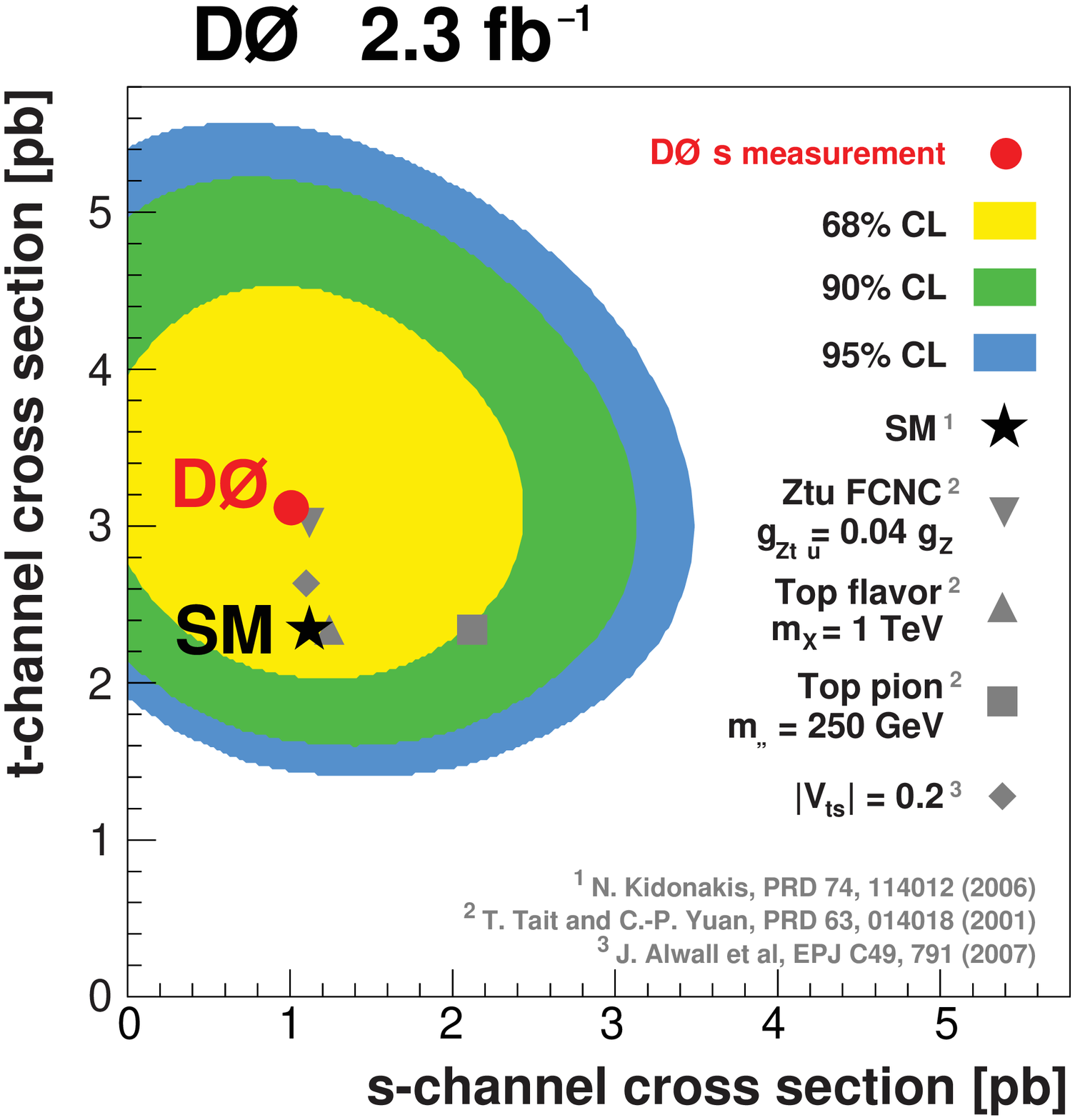}\hspace{0.2in}
\includegraphics[width=2.5in]{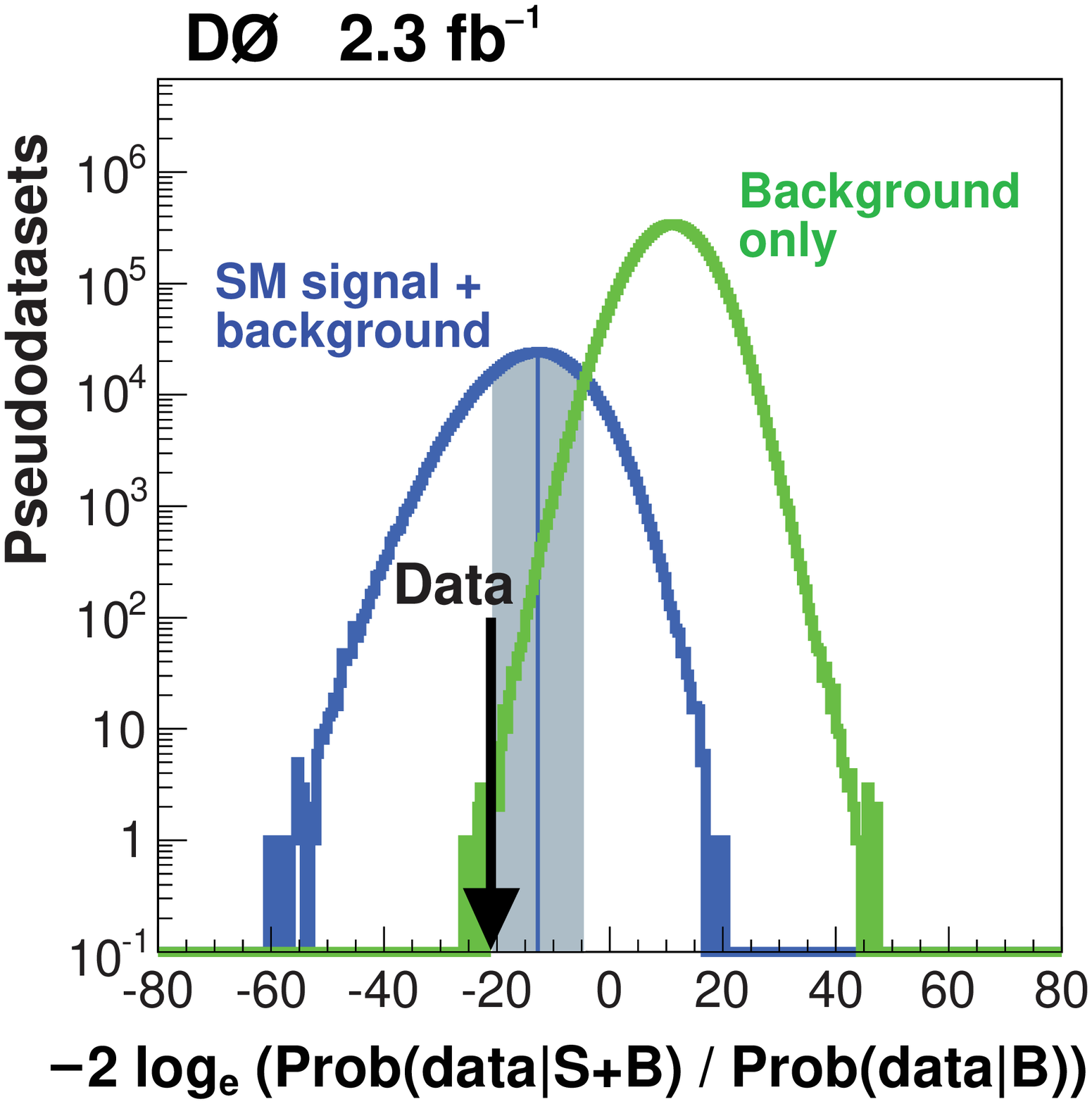}
\caption{Left plot: posterior probability density for t-channel and
s-channel single top quark production in contours of equal probability
density. Also shown are the measured cross sections, standard model
expectation, and several representative new physics
scenarios~\cite{newphysics}. Right plot: log-likelihood ratio plot
used to determine the significance of the t-channel measurement.}
\label{tchannel}
\end{center}
\end{figure}

\noindent The NLO theory predictions are~\cite{singletopxsec2}:
t-channel = 2.34~pb and s-channel = 1.12~pb, consistent with the
measurements. The recently published results~\cite{dzerotchannel} are
shown in Fig.~\ref{tchannel}.

\section{Single top quark production in the tau decay channel}

\begin{figure}[!h!tbp]
\begin{center}
\includegraphics[width=3.15in]{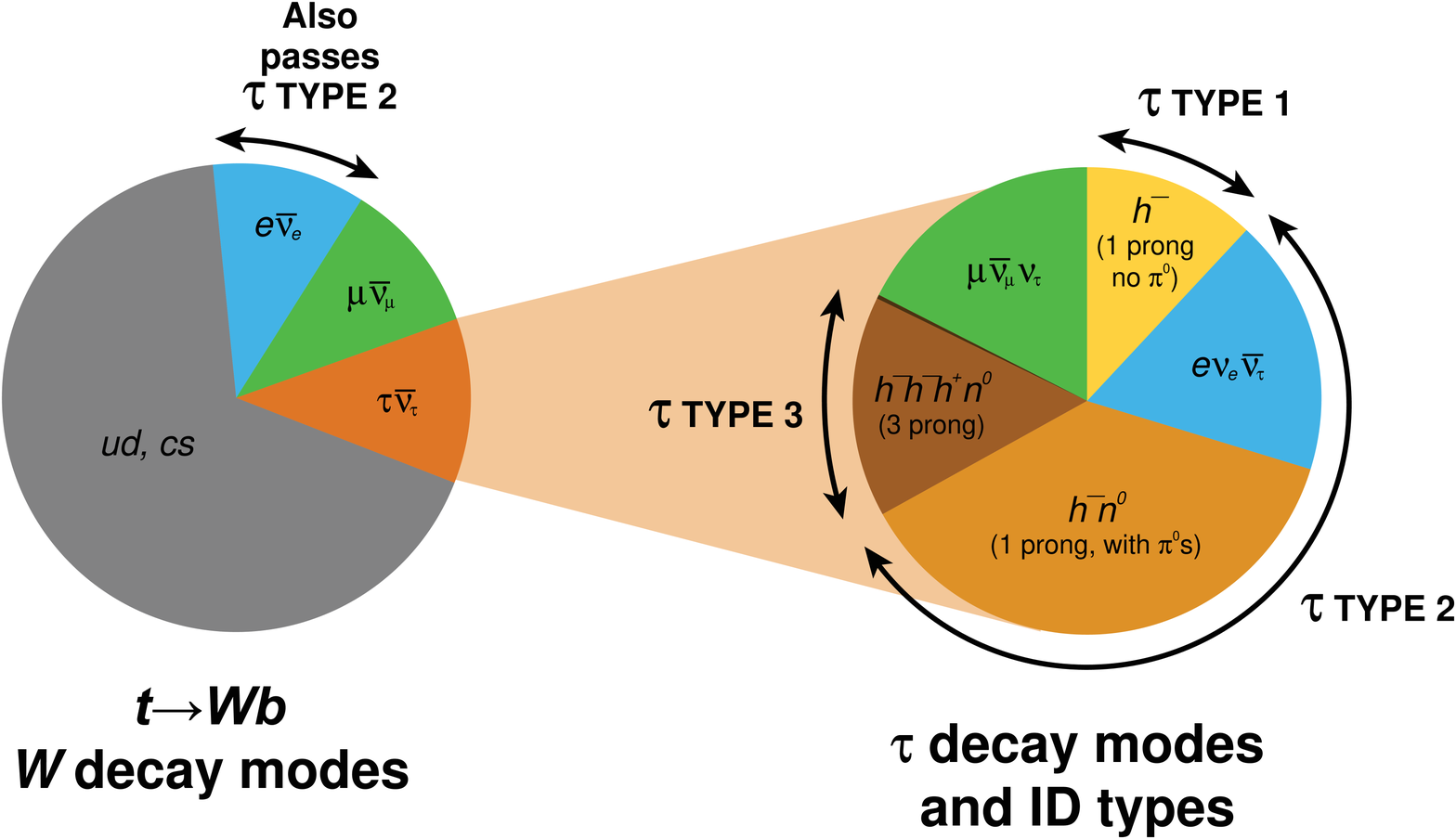}\hspace{0.1in}
\includegraphics[width=2in]{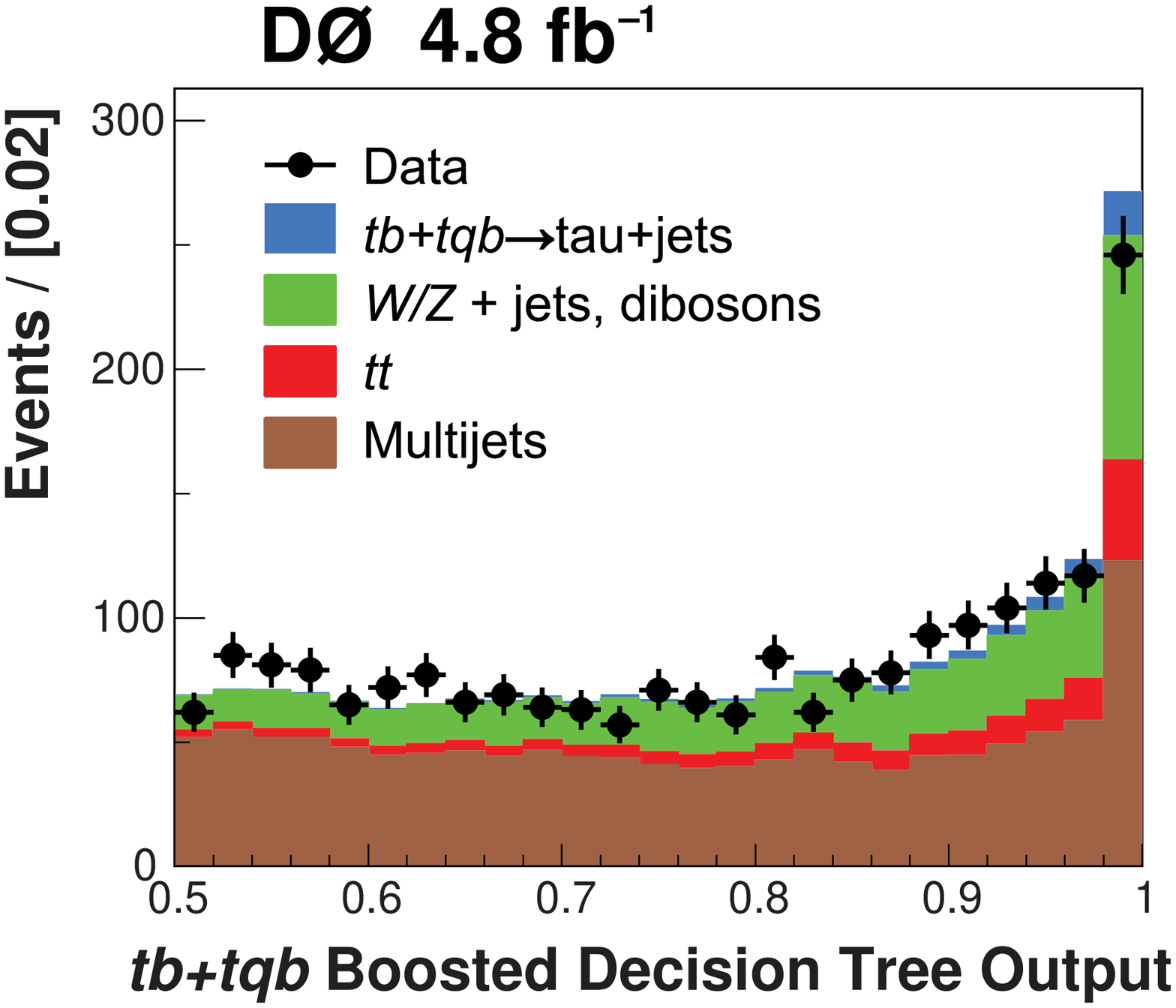}
\caption{Left image: tau identification by decay type. Right plot:
tau decay channel final boosted decision tree discriminant, with all
analysis channels combined.}
\label{taus}
\end{center}
\end{figure}

It is possible to add to the signal acceptance by searching in decay
channels not included in the main observation analysis. This can
provide additional sensitivity to reach 5$\sigma$ significance if
needed, as CDF has done with the inclusion of the {\met}+jets channel
in their observation result. {D\O} has performed a search for single
top quark production in the tau-lepton decay channel. That is, the top
quark decayed to a $W$~boson and a $b$~quark, the $W$~boson decayed to
a tau and a tau-neutrino, and the tau decayed hadronically to form a
narrow low-track-multiplicity jet. A new tau-identification algorithm
was developed for this search that is better tuned to find taus in
events containing additional jets (whereas {D\O}'s standard tau-ID is
optimized for taus in $Z \rar \tau \tau$ events, which are very
clean). Hadronically decaying taus were classified into three types
depending on the decay mode: Type~1 = calorimeter cluster + one track;
Type~2 = calorimeter cluster + one track + electromagnetic energy;
Type~3 = calorimeter cluster + two or three tracks. The algorithm used
boosted decision trees to discriminate tau jets from other jets,
gaining 8\%, 20\%, and 8\% in efficiency for Types 1, 2, and 3 taus
over the efficiencies obtained with the previous
neural-networked-based algorithm. The tau-ID efficiencies obtained are
76\%, 69\%, and 59\%, for 98\% background rejection. Taus that decayed
to electrons, and events with direct $W$~boson decay to electrons that
failed the main electron identification, were also selected by the
Type~2 algorithm in this analysis. The tau classifications are
illustrated in Fig.~\ref{taus}.

\vspace{-0.2in}
The analysis required one identified tau, missing transverse energy,
and two or three jets, with one or two of the jets identified as
having come from a $b$ decay. The dominant background, as one might
expect, was from multijet events where a jet was misidentified as a
tau. If the other jets were from light quarks or gluons, then
something had been misreconstructed to generated fake missing
transverse energy, and the $b$-tagged jets had fake tags. If there was
a $b\bar{b}$ pair in the event, then this could generate both real
{\met} and $b$~tags. Boosted decision trees were used in each
analysis channel to separate signal from background. A binned
likelihood calculation gave the following recently
published~\cite{dzerotauplb} cross section results for $m_t =
170$~GeV:

\vspace{-0.15in}
\begin{eqnarray*}
\sigma(p\bar{p}\rar tb+X, tqb+X) &=& 3.4^{+2.0}_{-1.8}
{\rm ~pb} {\rm ~~~(using~tau+jets~events)}\\
\sigma(p\bar{p}\rar tb+X, tqb+X) &=& 3.84^{+0.89}_{-0.83}
{\rm ~pb} {\rm~~~(all~channels~combined).}
\end{eqnarray*}


\section{Top quark width and lifetime}

In the standard model, the top quark's partial width is given at
leading order by:
\begin{eqnarray*}
\Gamma_{\rm LO}(t \rar Wb) = \frac{G_F m_t^3}{8 \pi \sqrt{2}}
\times |V_{tb}|^2.
\end{eqnarray*}
At next-to-leading order, and ignoring terms of order $m_b^2/m_t^2$,
$\alpha_s^2$, and $(\alpha_s/\pi)M_W^2/m_t^2$, the partial width
becomes~\cite{theorytopwidth}:
\begin{eqnarray*}
\Gamma_{\rm NLO}(t \rar Wb) & = &
\Gamma_{\rm LO}(t \rar Wb)
\left(1-\frac{M_W^2}{m_t^2}\right)^2
\left(1+2\frac{M_W^2}{m_t^2}\right)
\left[1-\frac{2\alpha_s}{3\pi}
\left(\frac{2\pi^2}{3}-\frac{5}{2}\right)\right]\\
& = & 1.26{\rm ~GeV~for~} m_t = 170{\rm ~GeV}.
\end{eqnarray*}

We have used our measured t-channel single top quark cross section to
determine the top quark partial width, since both are proportional to
$|V_{tb}|^2$ and hence proportional to each other, as proposed by
C.-P.~Yuan~\cite{yuan}. First, we removed the assumption of three
quark-generations from the t-channel measurement by scaling it by our
measurement of the top quark branching
fraction~\cite{dzerobranchingfrac}:
\begin{eqnarray*}
\sigma(p\bar{p}\rar tqb+X) \times {\cal{B}}(t\rar Wb)
& = & 3.14^{+0.94}_{-0.80}{\rm ~pb}\\
{\cal{B}}(t\rar Wb)
& = & 0.962 ^{+0.068}_{-0.066}({\rm stat})
^{+0.064}_{-0.052}({\rm syst})
\end{eqnarray*}
and then we scaled the measured cross section by the standard
model values for the partial width and the cross
section~\cite{singletopxsec2} at $m_t = 170$~GeV:
\begin{eqnarray*}
\Gamma(t\rar Wb)
& = & \sigma(p\bar{p}\rar tqb+X) \times
\frac{\Gamma_{\rm NLO}(t\rar Wb)}
{\sigma_{\rm NLO}(p\bar{p}\rar tqb+X)}\\
\sigma_{\rm NLO}(p\bar{p}\rar tqb+X)
& = & 2.14 \pm 0.18 {\rm ~pb.}
\end{eqnarray*}

The calculation was performed using a Bayesian technique, with all
statistical and systematic uncertainties and their correlations
included. The most probable value for the partial width and its
uncertainty, defined by the position of the peak of the
posterior and its width, is:
\begin{eqnarray*}
\Gamma(t\rar Wb) = 1.92^{+0.58}_{-0.51} {\rm ~GeV.}
\end{eqnarray*}

The total top quark width comes from the partial width using a similar
Bayesian calculation:
\begin{eqnarray*}
\Gamma(t)
= \frac{\Gamma(t\rar Wb)}{{\cal{B}}(t\rar Wb)}
= 1.99^{+0.69}_{-0.55} {\rm ~GeV.}
\end{eqnarray*}
This can be converted to the top quark lifetime using the reduced
Planck constant:
\begin{eqnarray*}
\tau(t)
= \frac{\hbar}{\Gamma(t)}
= \left(3.3^{+1.3}_{-0.9}\right) \times 10^{-25} {\rm ~s,}
\end{eqnarray*}
close to the standard model predicted value~\cite{theorytopwidth} of
$5.2 \times 10^{-25}$~s for $m_t = 170$~GeV. These results have
recently been submitted for publication~\cite{dzerotopwidth}.

\clearpage


\section{Flavor-changing neutral currents $tgq$}

\begin{figure}[!h!tbp]
\begin{center}
\includegraphics[width=5.25in]{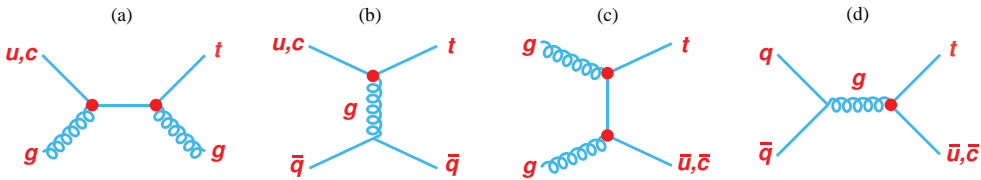}
\caption{Leading order Feynman diagrams for FCNC gluon coupling
between an up or a charm quark and a top quark. The red circles
indicate the effective FCNC coupling, possible at either of the two
vertices in (a) and (c) for which the amplitudes are properly summed.}
\label{feynman_fcnc}
\end{center}
\end{figure}

If flavor-changing neutral currents (FCNC) were to exist, then it
would be possible for the top quark to couple to a lighter up-type
quark and a gluon. This is illustrated in Fig.~\ref{feynman_fcnc}.
The process in Fig.~\ref{feynman_fcnc}(a) forms 83\% of the total rate
when the up-type quark is an up quark and 66\% when it is charm quark.

\begin{figure}[!h!tbp]
\begin{center}
\includegraphics[height=2.2in]{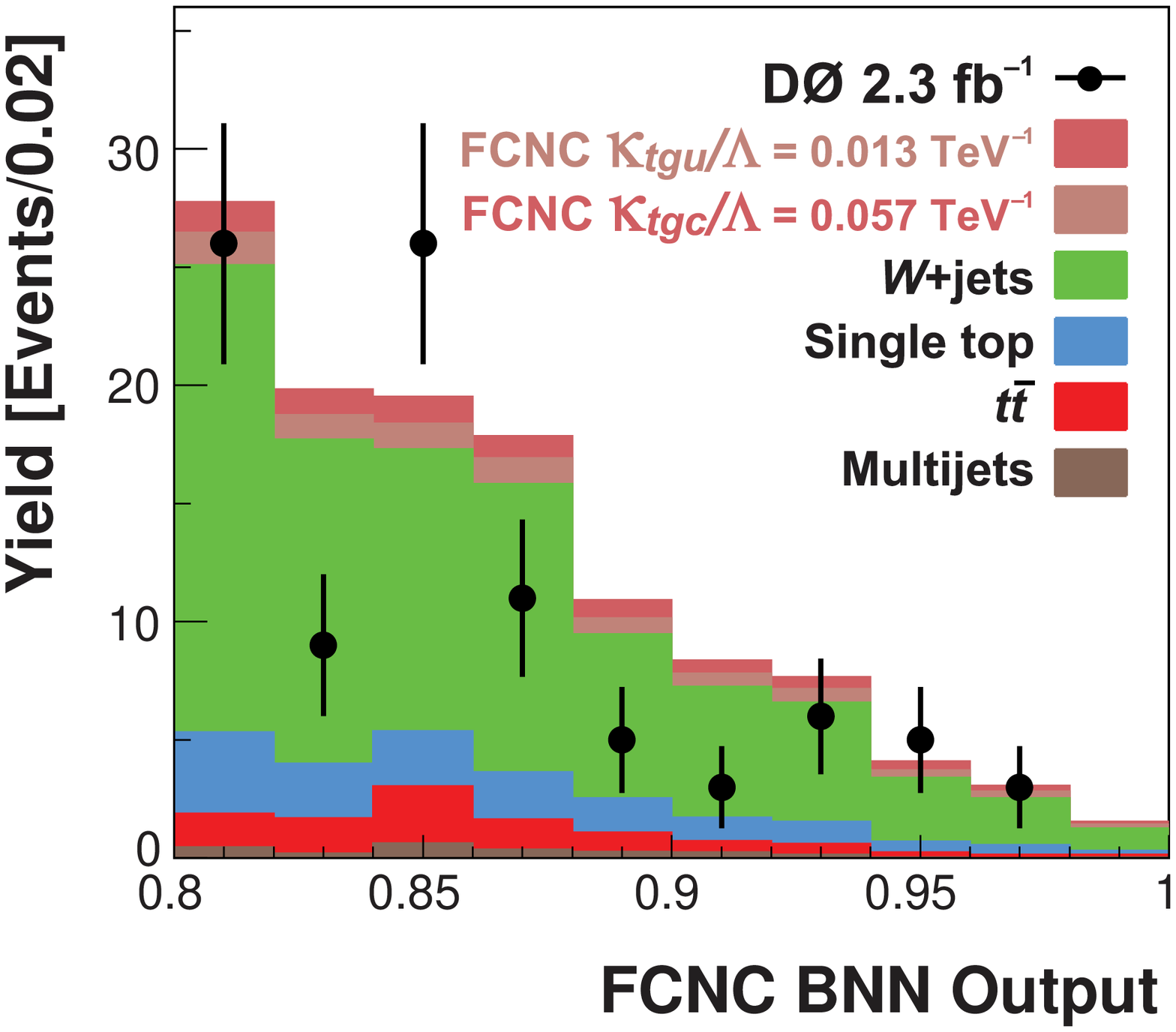}\hspace{0.2in}
\includegraphics[height=2.2in]{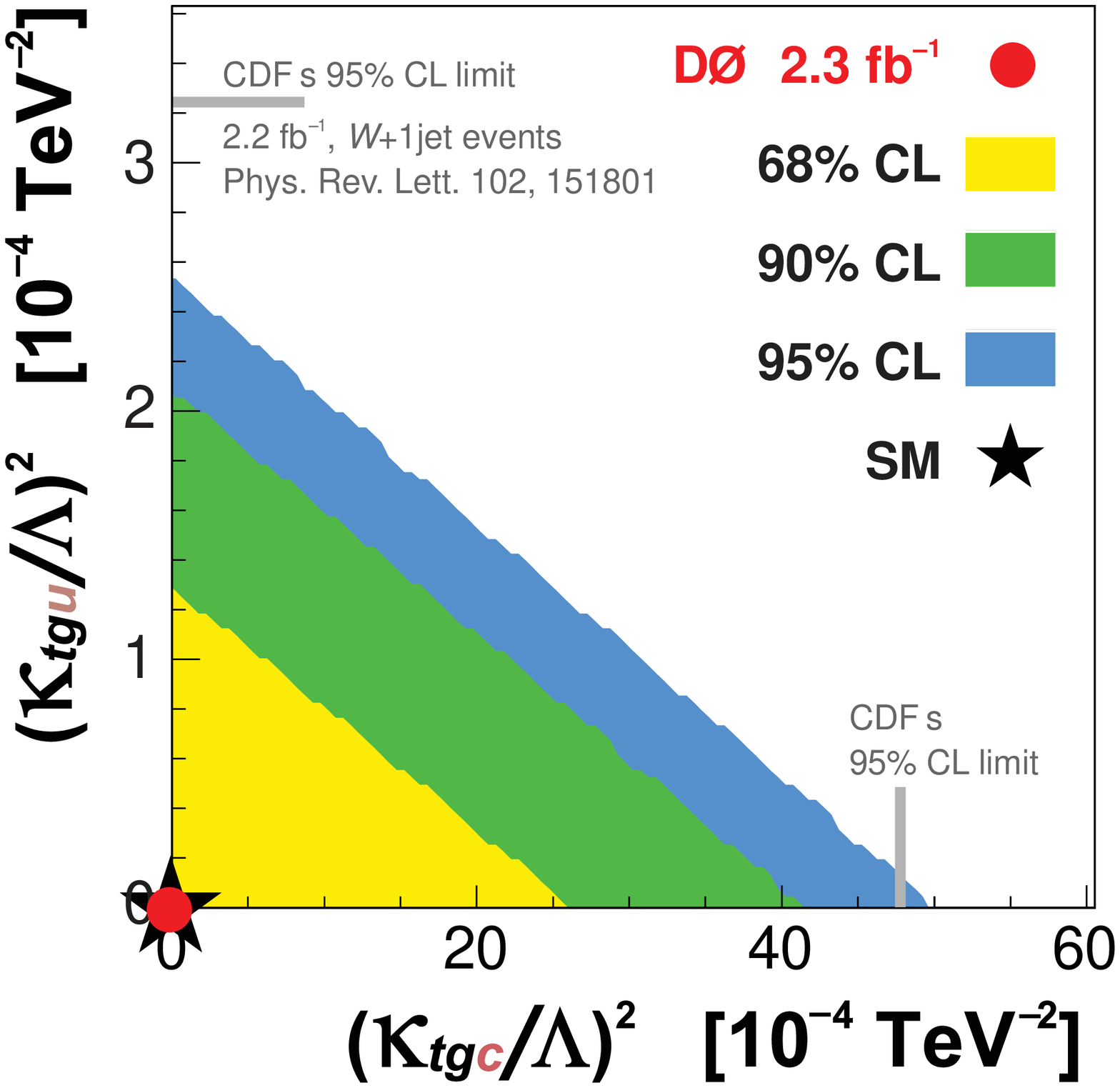}
\caption{Left plot: close-up of the Bayesian neural network signal
region. Right plot: Two-dimensional Bayesian posterior probability as
a function of the square of the couplings.}
\label{fcnc_plots}
\end{center}
\end{figure}

We have recently completed a search for the production of single top
quarks together with a light quark or gluon, instead of a bottom
quark. (Note that the simpler process $qg\rar t$ has only one jet in
the final state, from the top quark decay, so does not pass our
observation analysis requirement of at least two jets. The CDF
Collaboration used this production mode with one jet for their FCNC
analysis~\cite{cdffcncprl}.) For this new analysis, we used the
2.3~fb$^{-1}$ dataset and background model from the observation
analysis, which is ten times larger than used in our first
analysis~\cite{dzerooldfcnc}. The selection criteria were the same as
in the observation, except that we required exactly one $b$-tagged jet
to identify the $b$ from the top quark decay. Bayesian neural networks
(BNN) were used to separate signal from background with about 24
variables per channel, for a total of 54 variables. These include
individual object and event kinematics, top quark reconstruction, jet
widths, and angular correlations. We used a binned likelihood
calculation with the discriminant outputs to set limits on the
anomalous coupling constants, the signal cross sections, and the top
quark branching fraction to $gq$. The results are shown in
Table~\ref{fcnc_results} and illustrated in Fig.~\ref{fcnc_plots}.
They have been submitted for publication~\cite{dzeronewfcnc}.

\begin{table}[!h!tbp]
\caption[fcnc_results]{Observed 95\% C.L. upper one-dimensional
limits on the FCNC cross sections, couplings, and branching
fractions.}
\label{fcnc_results}
\begin{narrowtabular}{2cm}{lcc}
\hline
                       &        $tgu$         &        $tgc$        \\
\hline
Cross section          & 0.20 pb              & 0.27 pb             \\
$\kappa_{tgq}/\Lambda$ & 0.013 TeV$^{-1}$     & 0.057 TeV$^{-1}$    \\
${\cal{B}}(t\rar gq)$  & $2.0 \times 10^{-4}$ & $3.9 \times 10^{-3}$\\
\hline
\end{narrowtabular}
\end{table}

\clearpage

\section{Summary}

Since {D\O}'s observation of single top quark production in March
2009, we have produced four new publications using the same
dataset. We have measured t-channel $tqb$ production separately from
s-channel $tb$ production, with 4.8$\sigma$ significance for the
t-channel signal~\cite{dzerotchannel}. We developed a new more
efficient tau identification and measured the single top quark cross
section in the hadronic tau decay channel~\cite{dzerotauplb}. We used
the t-channel cross section measurement together with our value for
the branching fraction for top to decay to $Wb$ in top pair events to
determine the partial and total widths of the top quark and its
lifetime~\cite{dzerotopwidth}. And we have searched for single top
quark production via flavor-changing neutral currents with $tgq$
couplings~\cite{dzeronewfcnc}. The possibilities do not end here. We
recently completed a search for heavy $W^{\prime}$ resonant production
with decay to $tb$ that will be submitted for publication shortly, and
we expect to finish a new analysis of a dataset over twice as large as
the observation one in the near future. {D\O} should collect
10~fb$^{-1}$ of data by the end of September 2011, and there is a
possibility under discussion that the Tevatron will run for a further
three years to give an additional 7~fb$^{-1}$ for each experiment. We
are excited at the physics prospects in store for us.


\acknowledgments

I would like to thank the organizers of the workshop and other
attendees for an extremely interesting and enjoyable week. This work
is funded by a grant from the U.S. Department of Energy.

\clearpage


\end{document}